# Flight Testing of Latch Valve with Lightweight LV-Servo Direct Drive Mechanism

Hao-Che Huang[1], Chih-Shin Chang[1,2], Jui-Cheng Hsu[2] and Shih-Sin Wei[1,2,3]

**Institution:**

[1] Department of Mechanical Engineering, National Yang Ming Chiao Tung University, Taiwan

[2] Advanced Rocket Research Center, National Yang Ming Chiao Tung University, Taiwan

[3] Institude of Space System Engineering, National Yang Ming Chiao Tung University, Taiwan

**Corresponding author**

Address correspondence and reprint requests to Hao-Che Huang, Department of Mechanical Engineering, National Yang Ming Chiao Tung University, No. 1001, Daxue Rd. East Dist., Hsinchu City 300093, Taiwan

E-mail: robert4465.c@nycu.edu.tw

**Running title:** Flight Testing of Lightweight LV-Servo Latch Valve

**Word count:** 2150 words

# Abstract

In the field of rocket technology, the latch valve assumes a pivotal role in regulating the flow of fuel gases and liquids to ensure the requisite energy supply. This project endeavors to innovate by replacing the conventional step motor mechanism with a servo motor for latch valve control. The selected servo motor, boasting a more compact form factor and reduced mass, aligns seamlessly with the project's overarching objectives. While servo motors offer myriad advantages, it is imperative to acknowledge and address the constraints of their maximum output torque to guarantee the latch valve's reliable operation. Furthermore, as a rocket ascends, it encounters significant fluctuations in internal temperature and pressure. Consequently, rigorous environmental testing becomes paramount to validate the servo motor's performance under these dynamic conditions, thus ensuring the latch valve's unwavering functionality. The primary focus of this project is the design and testing of the mechanism's performance in simulated rocket environments, achieved through the implementation of the servo motor for latch valve control. The results reveal that the servo motor demonstrated its effectiveness and reliability in controlling the latch valve under the rigorous environmental conditions of rocket flight.

# 1. Introduction

In the field of rocket propulsion systems, precise control of fuel flow was crucial for optimal performance and mission success. Latch valves played a critical role in this process, regulating the flow of fuel gases and liquids. Traditionally, step motors had been employed to actuate these valves. However, recent advancements in servo motor technology offered potential advantages in terms of size, weight, and control precision [1, 2].

The required driving torque for the latch valve used in this project was approximately 30 kg-cm, as shown in Fig. 1. To meet this demand, the project employed the BMS-41A Servo Motor (Fig. 2) produced by Blue Bird Technology. According to Table 1, the BMS-41A offered a maximum output torque of 47.5 kg-cm, operated at a maximum voltage of 8.4V, and could withstand temperatures up to 85°C. For this project, the total voltage supplied to the LV-Servo Direct Drive Control Mechanism was set at 7.5V, with the servo motor receiving an operating voltage of approximately 6.75V, yielding an expected output torque of 38.1 kg-cm.

Given the extreme environmental conditions experienced during rocket flight, including significant temperature and pressure variations [3, 4], it was essential to evaluate the performance of this novel mechanism under simulated operational conditions. To this end, we conducted two critical environmental tests, "the Temperature Cycling Environmental Test" and "the Vacuum Environmental Test".

This paper presented a comprehensive analysis of the LV-Servo Direct Drive Control Mechanism, including its design, implementation, and performance under various environmental conditions. We aimed to contribute to the ongoing development of more efficient and reliable rocket propulsion systems by demonstrating the potential of servo motor technology in latch valve control applications.

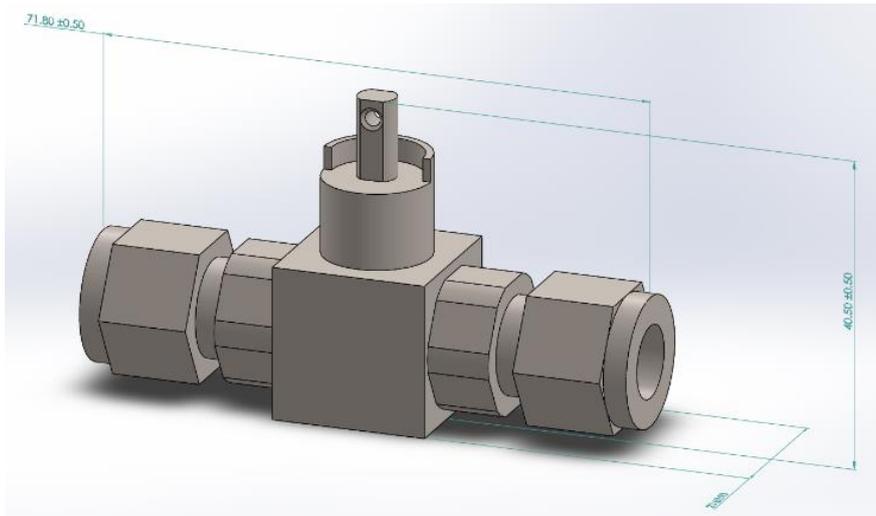

Figure 1. The latch valve

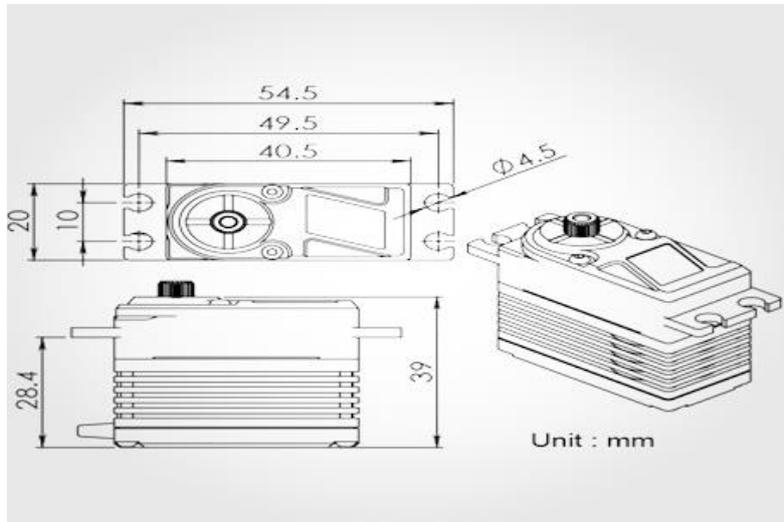

Figure 2. BMS-41A servo motor

Table 1. Specifications of the BMS-41A Servo Motor by Blue Bird Technology

| Model | BMS-41A |
|---|---|
| Max Output Torque (kg-cm) | 47.5 |
| Operating Voltage (V) | 4.5V ~ 8.5V |
| Dimensions (mm³) | 40.5 × 20 × 39 |
| Response Speed (sec/60°) | 0.11 |
| Control System | Digital Controller |
| Weight (g) | 74 ± 1 |
| Temperature Protection | > 85°C |

## 2. Research Methods

The LV-Servo Direct Drive Control Mechanism was composed of two primary components: the mechanical structure that connects the servo motor to the latch valve, and the microcontroller system. Additionally, the LV-Servo Direct Drive Control Mechanism underwent the thermal cycling environment test and the vacuum environment test to ensure reliable operation within the rocket vehicle.

**Mechanical Structure**

The design of the mechanical structure connecting the servo motor to the latch valve is illustrated in Fig. 3(a). It consisted of the housing (Fig. 3(b)), the primary connecting component (Fig. 3(c)), and the secondary connecting component (Fig. 3(d)). These parts worked together to transmit the torque from the servo motor to the latch valve, enabling its operation.

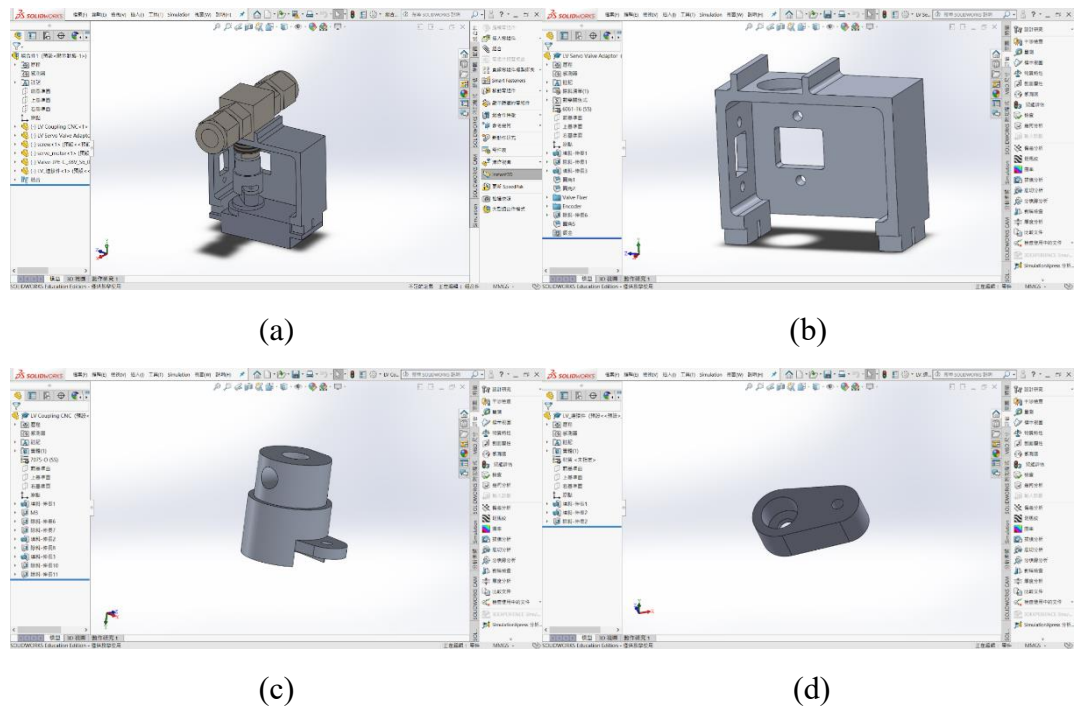

(a)                  (b)

(c)                  (d)

Figure 3. Mechanical Structure of LV-Servo Direct Drive Control Mechanism: (a) the mechanical structure, (b) the housing, (c) the primary connecting component and (d) the secondary connecting component

**Microcontroller System**

The microcontroller system comprised an Arduino Uno, an SD Card Module, a Voltage Current Sensor (MAX471), and a Power Supply (Fig. 4). The Arduino Uno output angle signals to the servo motor, causing it to change its angle and thereby open or close the latch valve (Fig. 5). Simultaneously, the Voltage Current Sensor monitored the servo motor's operating voltage and current, recording this data onto the SD card. The Arduino code used for this process is listed in Appendix.

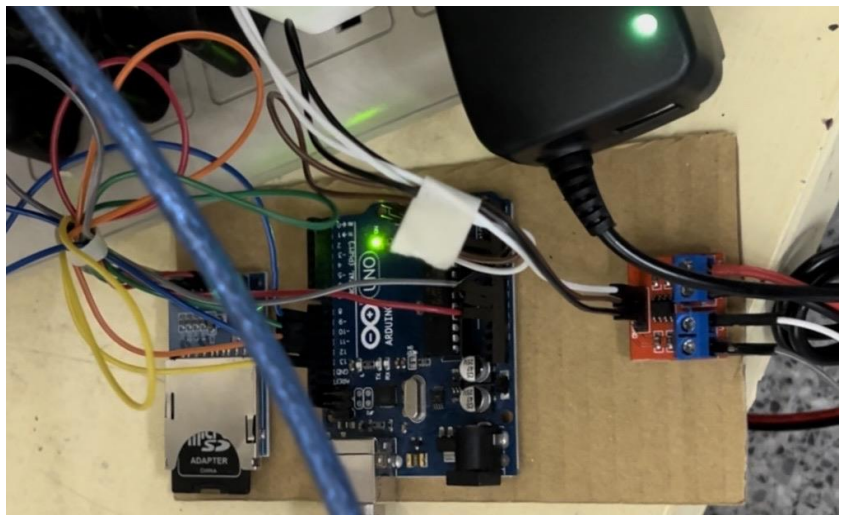

Figure 4. Arduino Uno, SD card module, voltage current sensor and power supply

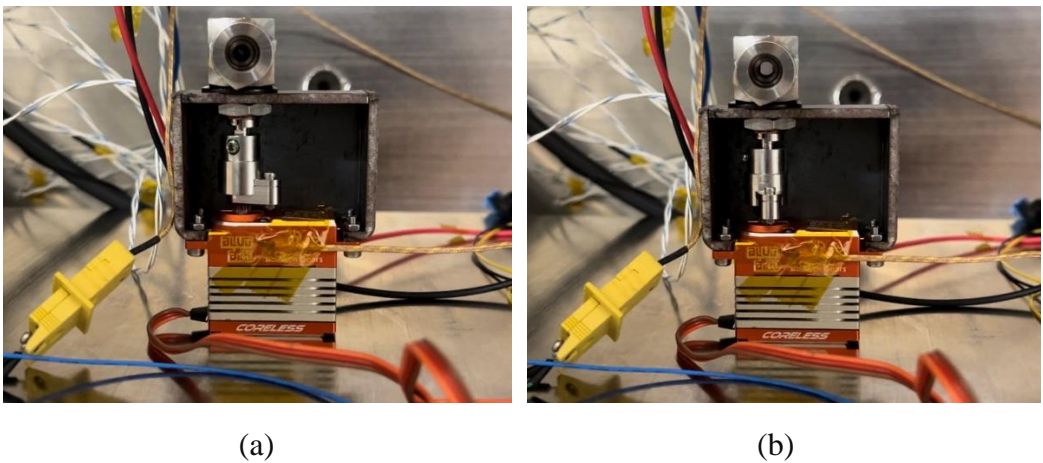

(a)                        (b)

Figure 5. LV-Servo direct drive control mechanism: (a) Latch valve close and (b) Latch valve open

**Environmental Test – Temperature Cycling**

The objective of this test was to verify whether the LV-Servo Direct Drive Control Mechanism can endure the environmental temperatures experienced during the flight of an arrow and operate normally. The experimental procedure (Fig. 6) included two thermal cycling tests (Fig. 7). And the setup of the temperature cycling environmental test is shown in Fig. 8.

**a. Test Items:**

LV-Servo Direct Drive Control Mechanism

**b. Test Equipment:**

Thermal Cycling Test Chamber

**c. Thermal Cycling Profile:**

One cycle consists of:

- Holding at room temperature (5 minutes)
- Heating from room temperature to 60℃ at a rate of 2℃ per minute
- Holding at 60°C (30 minutes)
- Cooling from 60°C to -15°C at a rate of 1℃ per minute
- Holding at -15°C (30 minutes)
- Heating from -15°C to room temperature at a rate of 2℃ per minute

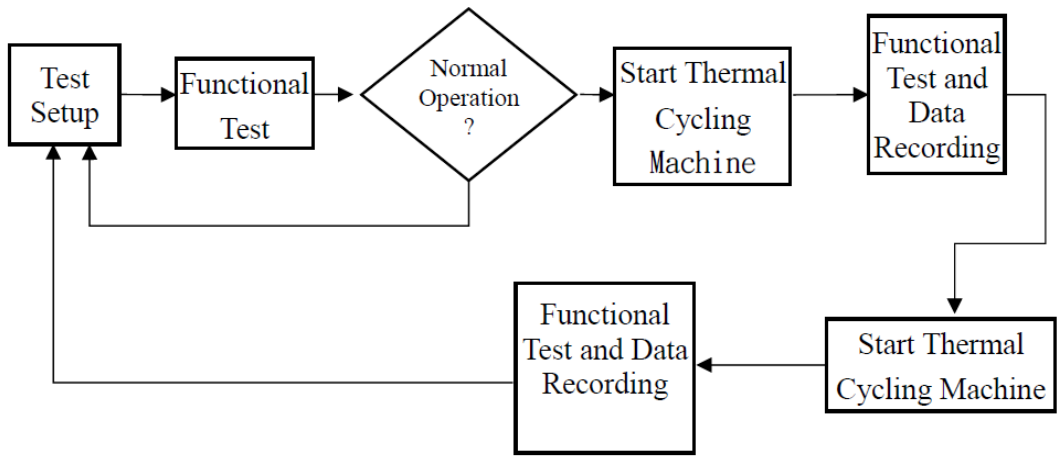

Figure 6. Procedure diagram of temperature cycling

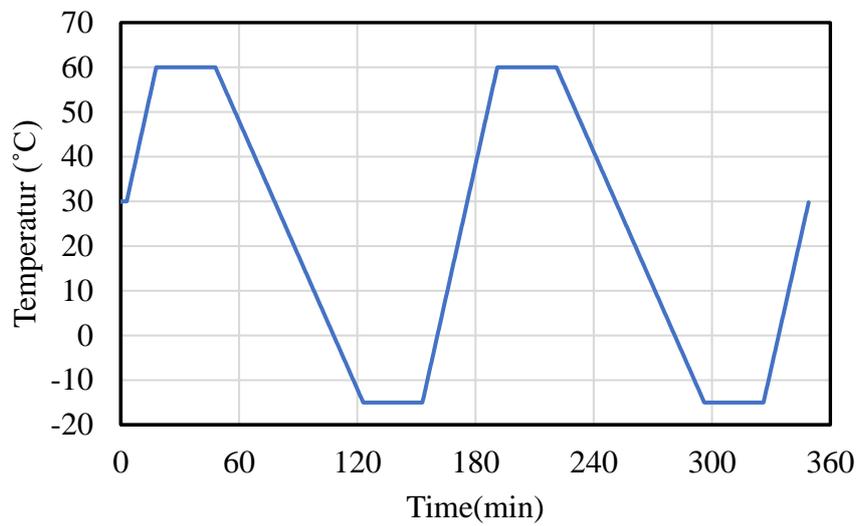

Figure 7. Temperature control curve of thermal cycling machine

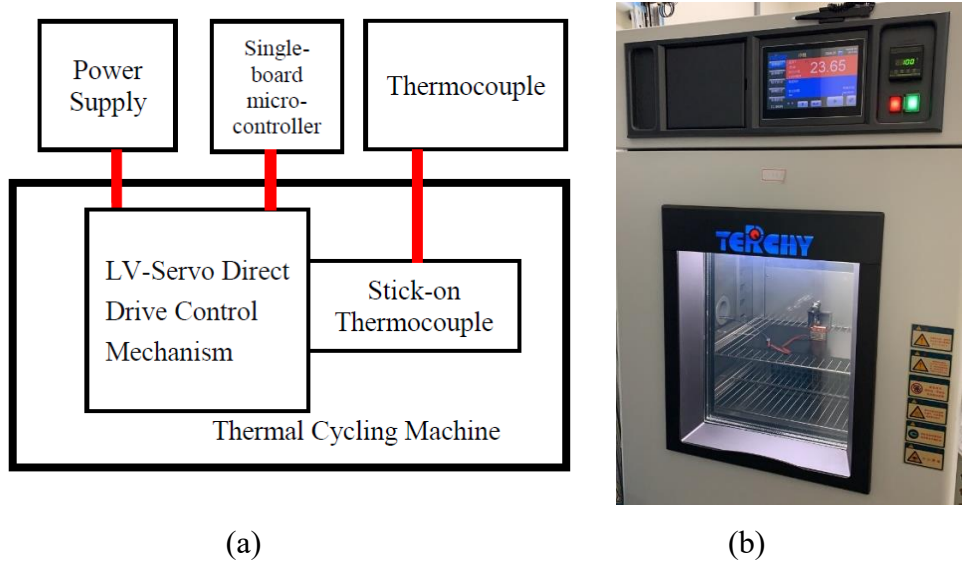

(a)                  (b)

Figure 8. Temperature cycling environmental test: (a) Test setup diagram and (b) Test setup photo

**Environmental Test – Vacuum**

The objective of this test was to verify whether the LV-Servo Direct Drive Control Mechanism can withstand the vacuum environment experienced during the flight of an arrow and operate normally. This experiment was conducted twice, once with the vacuum chamber and LV-Servo Direct Drive Control Mechanism uninsulated, and once with them insulated. Figure 9 shows the experimental procedure which included one vacuum test. Figures 10 shows the setup of the vacuum environmental test.

a. **Test Items:** LV-Servo Direct Drive Control Mechanism

b. **Test Equipment:** Small Vacuum Chamber

c. **Vacuum Cycling Profile:** The vacuum degree requirement is that the vacuum chamber pressure must be $\leq 1 \times 10^{-3}$ Torr, maintaining this vacuum degree for at least 30 minutes.

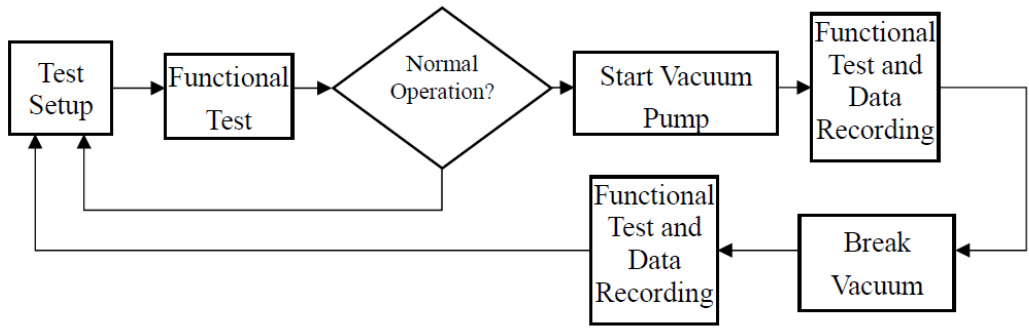

Figure 9. Procedure diagram of temperature cycling

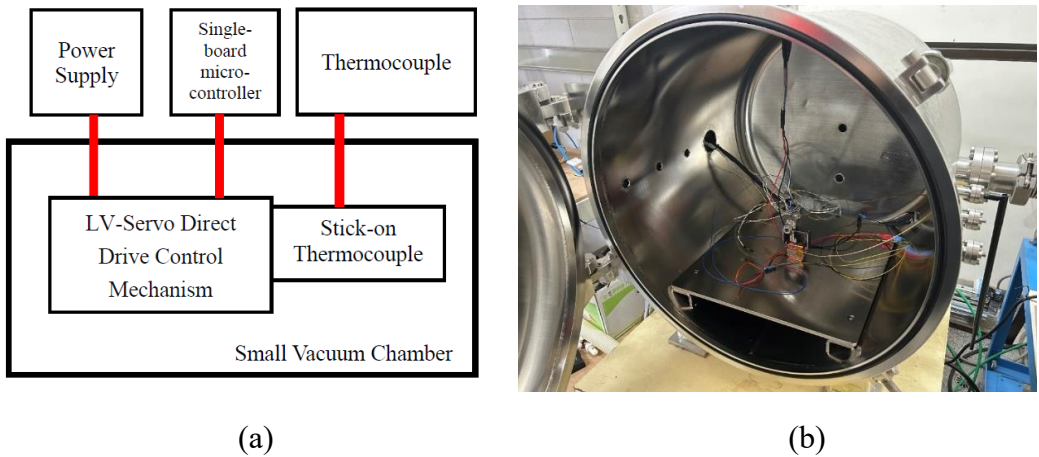

(a)                                      (b)

Figure 10. Vacuum environmental test: (a) Test setup diagram and (b) Test setup photo

# 3. Results and Discussion

The LV-Servo Direct Drive Control Mechanism demonstrated robust performance throughout both the Temperature Cycling Environmental Test and the Vacuum Environmental Test. The mechanism operated effectively within a temperature range of -15℃ to 60℃ and under vacuum conditions of $1.6\times10^{-4}$ Torr (denoted as $p_1$), ensuring the latch valve could open and close as intended under various environmental conditions.

The monitoring of the BMS-41A servo motor using the MAX471 Voltage Current Sensor demonstrated significant variations in operational characteristics when subjected to different environmental conditions, with specific voltage, current, and power values observed.

Under standard conditions (760Torr, 28.3°C), the servo motor exhibited a stable performance across different operational states, as shown in Figs. 11. During the idle phase, the motor maintained a voltage of 6.75V, with a current of 120mA, resulting in a power consumption of 0.81W. When the valve was being opened, the voltage slightly decreased to 6.45V, while the current increased to 400mA, leading to a power consumption of 2.6W. Similarly, during the valve closing phase, the voltage further decreased to 6.41V, with the current increasing to 435mA, and the power consumption reaching 2.8W.

In contrast, when the servo motor operated under vacuum conditions ($p_1$), there was a notable increase in current draw and power consumption during dynamic operations, as shown in Figs. 12. While the idle state remained consistent with the standard conditions, the valve opening phase saw the current rise to 435mA, with a corresponding power consumption of 2.8W at a voltage of 6.45V. During the valve closing phase, the current further increased to 470mA, with the voltage at 6.47V, resulting in a power consumption of 3.1W. These results suggest that the vacuum environment imposed additional load on the motor during the dynamic phases of operation.

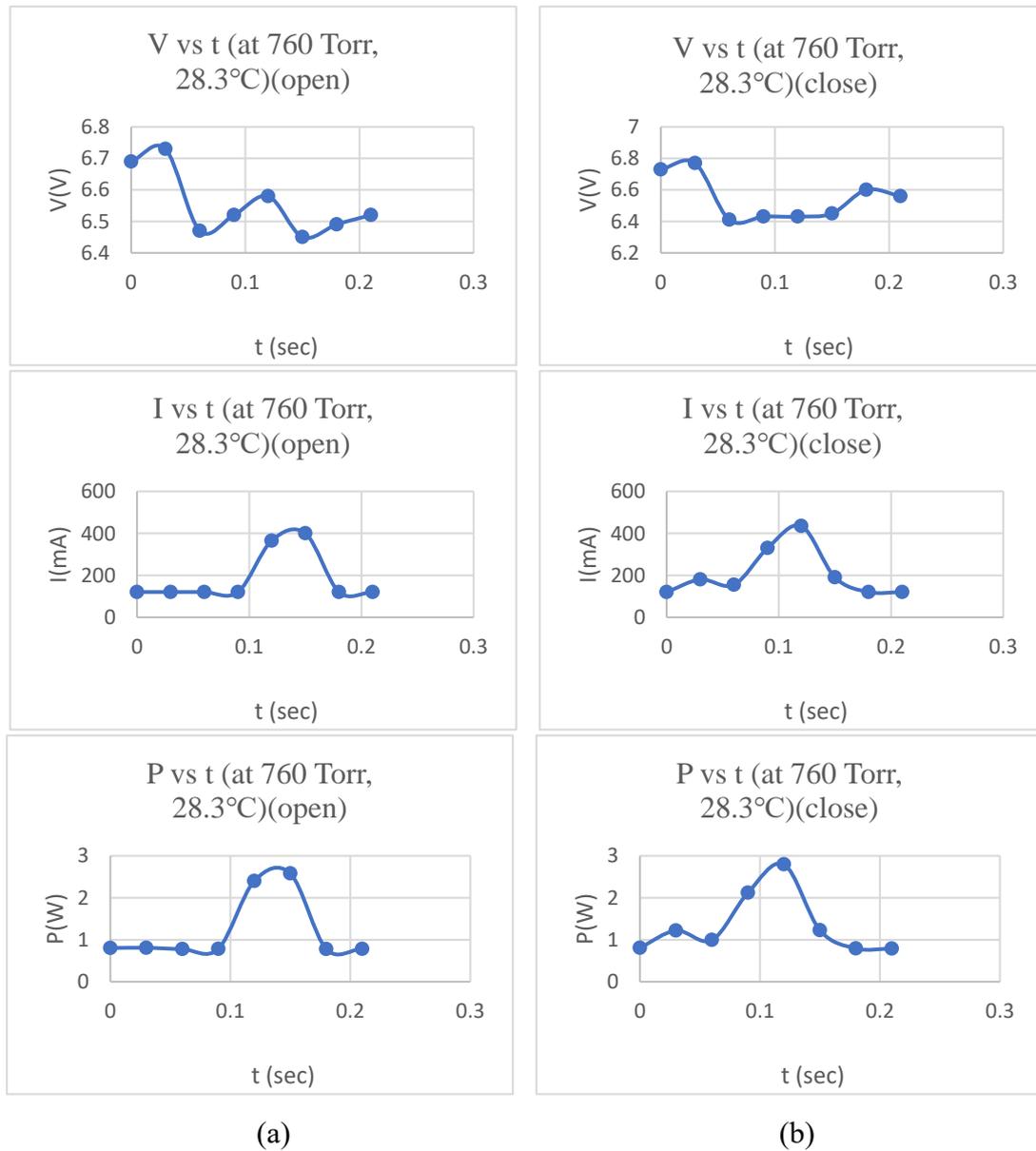

(a)          (b)

Figure 11. Voltage, Current, Power of Servo Motor at 760Torr, 28.3°C, Uninsulated (a) LV open (b) LV close

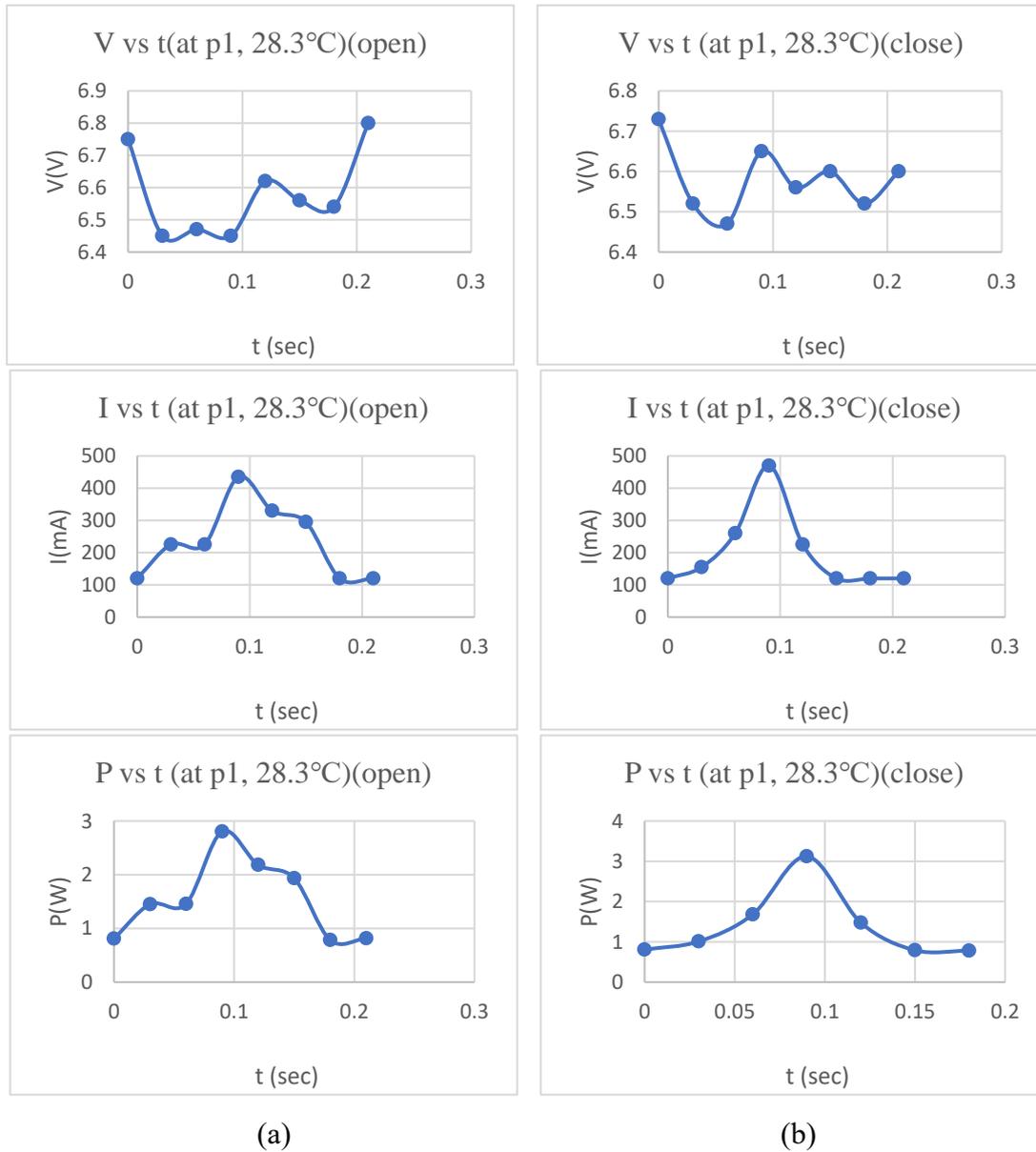

(a)                  (b)

Figure 12. Voltage, Current, Power of Servo Motor at $p_1$, 28.3°C, Uninsulated (a) LV open (b) LV close

In a more isolated vacuum chamber with a slightly elevated temperature ($p_1$, 31.7°C), the servo motor exhibited even more pronounced changes, as shown in Figs. 13. During the idle phase, the motor's performance remained consistent at 6.75V, 120mA, and 0.81W. However, during the valve opening phase, the current surged to 630mA, leading to a higher power consumption of 4W at a voltage of 6.43V.

Conversely, during the valve closing phase, the current dropped to 330mA, with a voltage of 6.47V, resulting in a reduced power consumption of 2.2W.

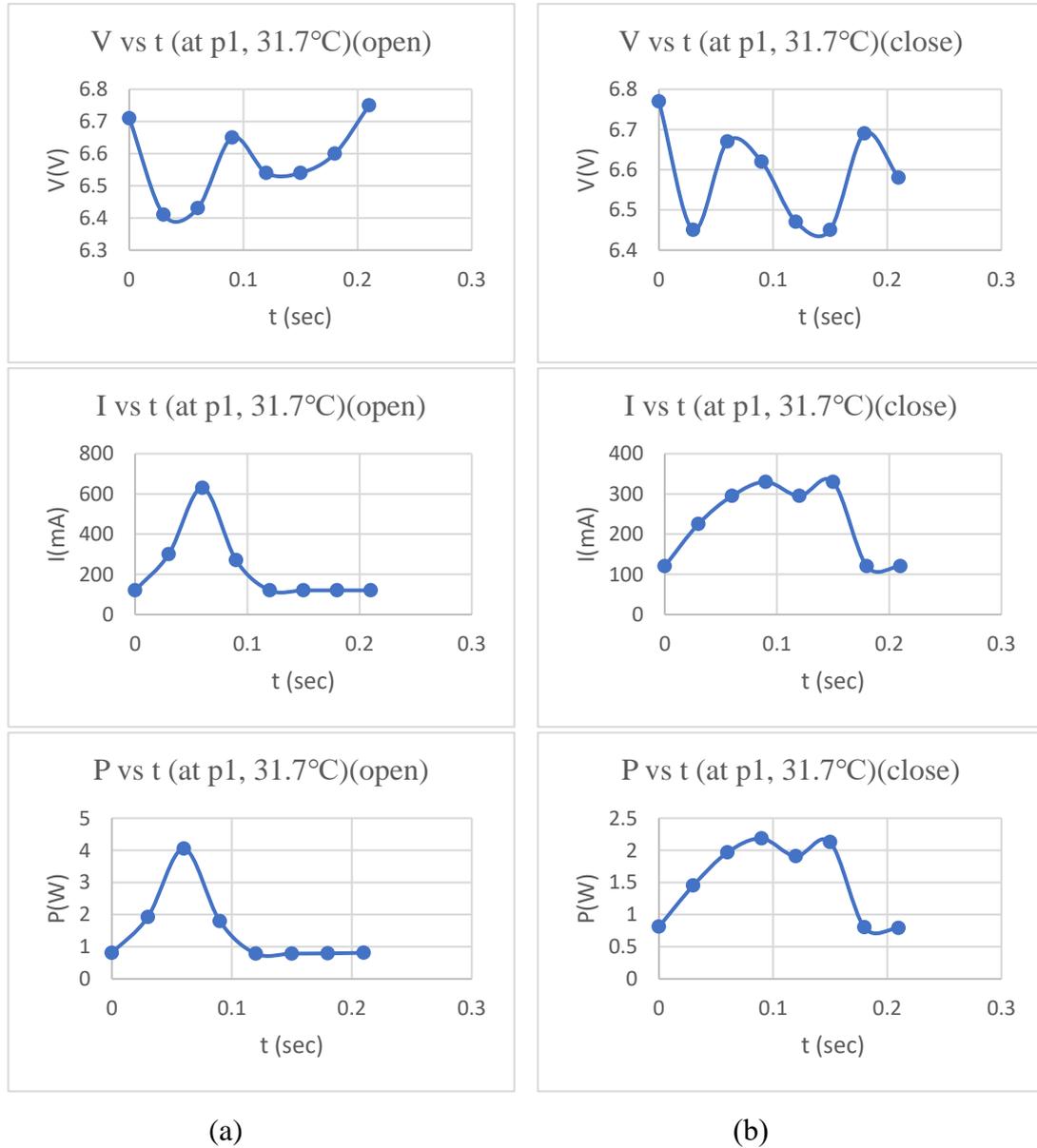

(a)          (b)

Figure 13. Voltage, Current, Power of Servo Motor at $p_1$, 31.7°C, insulated (a) LV open (b) LV close

The results indicated that vacuum conditions had minimal impact on the idle state of the servo motor. However, during active operations, there was a notable increase in current draw and power consumption, particularly when the vacuum chamber was isolated from the mechanism.

Additionally, Heat accumulation was observed when the vacuum chamber was isolated, resulting in a temperature increase of approximately 3.4℃ in the BMS-41A. Within 30 minutes in the $p_1$ environment, the mechanism experienced a temperature increase of 0.6℃ (Fig. 14).

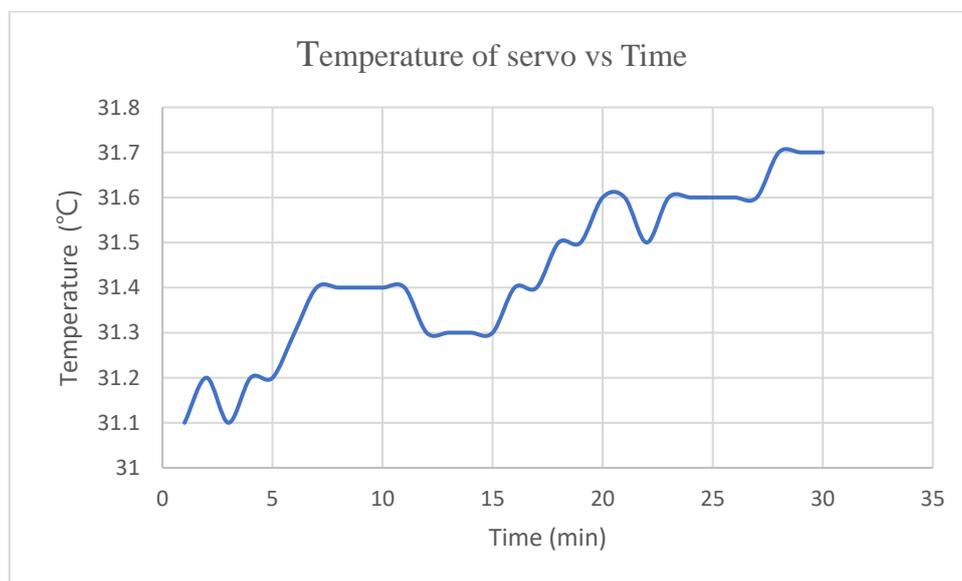

Figure 14. Temperature of LV-Servo Direct Drive Control Mechanism at pressure $p_1$

We also found that the standby power consumption remained consistent at approximately 0.81W across all conditions. However, under vacuum conditions with heat accumulation, the working power increased slightly during valve operations. In one instance, the operating current rose to approximately 630mA with a power consumption of about 4W when opening the valve. This increase was hypothesized to be due to reduced motor efficiency caused by heat accumulation, necessitating higher current to compensate for reduced torque. Despite these variations, the average power consumption for most valve operations remained around 3W, demonstrating the mechanism's ability to maintain relatively stable performance across different environmental conditions.

In conclusion, the LV-Servo Direct Drive Control Mechanism exhibited reliable performance under both the Temperature Cycling Environmental Test and the Vacuum Environmental Test. Its ability to maintain normal operation within the specified temperature range and vacuum conditions confirmed its suitability for aerospace applications, highlighting its robustness and adaptability in the demanding environment of rocket propulsion systems.

# 4. Conclusion

The LV-Servo Direct Drive Control Mechanism demonstrated its effectiveness and reliability in controlling the latch valve under the rigorous environmental conditions of rocket flight. The mechanism maintained stable performance across both the thermal cycling and vacuum environment tests, confirming its ability to operate within temperature ranges from -15°C to 60°C and under vacuum conditions down to $1\times10^{-4}$ Torr. The BMS-41A servo motor exhibited consistent idle power consumption and manageable variations in current draw and power consumption during dynamic valve operations, particularly under vacuum and heat accumulation conditions.

These results indicate that the LV-Servo Direct Drive Control Mechanism is well-suited for aerospace applications, particularly in lightweight and high-efficiency propulsion systems. The robustness and adaptability of the system in handling fluctuating temperatures and pressure conditions make it a viable option for use in rocket vehicles.

# Appendix

*Code for the LV-Servo Direct Drive Control Mechanism:*

```
servo_detection_SDmodule.ino

 1
 2
 3   /* Connection pins:
 4
 5   Arduino      Current Sensor B43
 6   A0                VT
 7   A1                AT
 8   +5V               VIN
 9   +5V               VOUT
10   GND               GND
11   */
12
13   #include <SPI.h>
14   #include <SD.h>
15   #include "SdFat.h"
16
17   #define SD_CS_PIN SS   // CS腳位預設是SS腳位10,如果有多個SPI設備,要設定不同的腳位
18   File myFile;
19
20   #include <Wire.h>
21
22   #define VT_PIN A0 // connect VT
23   #define AT_PIN A1// connect AT
24
25   #define ARDUINO_WORK_VOLTAGE 5.0
```

```
servo_detection_SDmodule.ino

26
27   #include <Servo.h>
28   Servo myservo;   // 定義servo對象,最多八個
29   int pos = 1050;
30   void setup()
31   {
32     Serial.begin(9600);
33     Serial.println("\n");
34     Serial.print("Initializing SD card...");
35
36       //判斷SD模組初始化是否成功
37       if (!SD.begin(SD_CS_PIN)) {
38         Serial.println("initialization failed!");
39         return;
40       }
41       Serial.println("initialization done.");
42
43       //打開一個檔名為test.txt的檔案,模式為寫入(FILE_WRITE)
44       //若檔案不存在,就會自動建議一個新的檔案
45       myFile = SD.open("test.txt", FILE_WRITE);
46
47       // 如果成功打開,就寫入文字
48       if (myFile) {
49         Serial.print("Writing to test.txt...");
50         myFile.println("start");
```

```
servo_detection_SDmodule.ino

51
52
53       Serial.println("done1.");
54     } else {
55       // 如果無法開啟檔案，就在監控視窗顯示訊息
56       Serial.println("error opening test.txt");
57     }
58
59   myFile.println("Voltage (V) / Current (A)");
60   myFile.close();
61   myservo.attach(9);   // 設置servo控制腳位
62   myservo.write(1050);      //回到15度
63   detect(analogRead(VT_PIN),analogRead(AT_PIN));
64   Serial.println("done2.");
65
66   delay(30000);
67   myFile = SD.open("test.txt", FILE_WRITE);
68   myFile.println("after 30s");
69   myFile.close();
70
71   detect(analogRead(VT_PIN),analogRead(AT_PIN));
72
73
74   delay(30000);
75   myFile = SD.open("test.txt", FILE_WRITE);
```

```
servo_detection_SDmodule.ino

76   myFile.println("after 30s\n");
77   myFile.close();
78   detect(analogRead(VT_PIN),analogRead(AT_PIN));
79   // 關閉檔案
80
81 }
82
83 void loop()
84 {
85   myFile = SD.open("test.txt", FILE_WRITE);
86   myFile.println(" one circuit finished.");
87   myFile.close();
88   detect(analogRead(VT_PIN),analogRead(AT_PIN));
89
90   for(pos = 1050; pos < 1950; pos += 150)
91   {
92     myservo.write(pos); delay(35);
93
94     detect(analogRead(VT_PIN),analogRead(AT_PIN));
95
96
97   }
98
99   detect(analogRead(VT_PIN),analogRead(AT_PIN));
100
```

```
servo_detection_SDmodule.ino                                                                                        ...
101     delay(30000);
102     myFile = SD.open("test.txt", FILE_WRITE);
103     myFile.println("after 30s");
104     myFile.close();
105     detect(analogRead(VT_PIN),analogRead(AT_PIN));
106
107     delay(30000);
108
109
110     myFile = SD.open("test.txt", FILE_WRITE);
111     myFile.println("after 30s\n");
112     myFile.close();
113     detect(analogRead(VT_PIN),analogRead(AT_PIN));
114
115     for(pos = 1950; pos > 1050; pos-=150)
116         {
117           myservo.write(pos);
118           delay(35);
119
120           detect(analogRead(VT_PIN),analogRead(AT_PIN));
121
122
123         }
124
125     detect(analogRead(VT_PIN),analogRead(AT_PIN));
```

```
servo_detection_SDmodule.ino                                                                                        ...
124
125     detect(analogRead(VT_PIN),analogRead(AT_PIN));
126
127     delay(30000);
128     myFile = SD.open("test.txt", FILE_WRITE);
129     myFile.println("after 30s");
130     myFile.close();
131
132     detect(analogRead(VT_PIN),analogRead(AT_PIN));
133
134     delay(30000);
135
136     Serial.println("done one circuit.");
137
138 }
139 void detect(int a, int b){
140     int vt_temp = a;
141     int at_temp = b;
142
143     double voltage = vt_temp * (ARDUINO_WORK_VOLTAGE / 1023.0) * 5*0.88;
144     double current = (-1)*(at_temp * (ARDUINO_WORK_VOLTAGE / 1023.0)/2.6-1.92)+0.12;
145     myFile = SD.open("test.txt", FILE_WRITE);
146     myFile.print("V: ");myFile.print(voltage); myFile.print(" _ "); myFile.print("A: "); myFile.println(current);
147     myFile.close();
148
149 }
```